\documentclass[usenatbib]{mnras}

\usepackage{natbib}
\usepackage{aas_macros,amssymb}
\usepackage{graphicx}
\begin{document}

\title[Cosmological Density Distribution]{Precision Prediction for the Cosmological Density Distribution}

\author[A. Repp \& I. Szapudi]{Andrew Repp\ \& Istv\'an Szapudi\\Institute for Astronomy, University of Hawaii, 2680 Woodlawn Drive, Honolulu, HI 96822, USA}

\date{MNRAS 473, 3598--3607 (2018)}

\label{firstpage}
\pagerange{\pageref{firstpage}--\pageref{lastpage}}
\maketitle

\begin{abstract}
The distribution of matter in the universe is approximately lognormal, and one can improve this approximation by characterizing the third moment (skewness) of the log density field. Thus, using Millennium Simulation phenomenology and building on previous work, we present analytic fits for the mean, variance, and skewness of the log density field $A$, allowing prediction of these moments given a set of cosmological parameter values. We further show that a Generalized Extreme Value (GEV) distribution accurately models $A$; we submit that this GEV behavior is the result of strong intrapixel correlations, without which the smoothed distribution would tend toward a Gaussian (by the Central Limit Theorem). Our GEV model (with the predicted values of the first three moments) yields cumulative distribution functions accurate to within 1.7 per cent for near-concordance cosmologies, over a range of redshifts and smoothing scales.
\end{abstract}

\section{Introduction}
The fluctuations in the cosmic microwave background (CMB) are remarkably Gaussian, implying a similarly Gaussian matter distribution at decoupling. It is the subsequent non-linear action of gravity that has produced the highly non-Gaussian distribution observable today \citep{FryPeebles1978, SSB1992, Gaztanaga1994}.

This distribution is approximately lognormal, a fact which emerges naturally under the assumption that peculiar velocities grow in linear fashion \citep{ColesJones}. Lognormality is however only a rough first-order approximation (see Fig.~\ref{fig:A_dist}), and in the framework of high-precision cosmology it is important to better characterize this distribution.

One reason for doing so is that forecasts of future surveys' effectiveness require assumptions about the distribution of the underlying matter field -- and a standard assumption is that this field is Gaussian (e.g. \citealp{Wang2010}). This assumption can result in significant overestimates obtainable from a survey -- up to an order of magnitude for amplitude-like parameters \citep{Repp2015}.

Accurate characterization of the matter distribution is essential to recovery of this information. It is the non-Gaussianity of the overdensity field that causes a sizable fraction of the information to escape its power spectrum \citep{RimesHamilton2005, NeyrinckSzapudi2007, Carron2011, CarronNeyrinck2012, Wolk2013}. By considering instead an ``optimal observable'' \citep{CarronSzapudi2013}, one can recapture this otherwise-lost information. However, to predict the power spectrum of the optimal observable for galaxy surveys (denoted $A^*$ -- see \citealp{CarronSzapudi2014}), one must know the underlying dark matter distribution (Repp \& Szapudi, submitted; see Section~\ref{sec:disc}). Furthermore, knowledge of the matter distribution allows determination of the galaxy bias function itself \citep{SzapudiPan2004}. Hence, to access all of the Fisher information in galaxy surveys, it is necessary to precisely describe the cosmological density distribution.

Various fits to the matter distribution have appeared in recent literature. \citet{Lee2017} use multiple Generalized Extreme Value (GEV) distributions to fit the density at various scales and redshifts. \citet{Shin2017} use a generalized normal distribution to describe the density for five cosmological models. \citet{Klypin2017} fit the matter distribution using a power law exponentially suppressed at both ends, considering two values of $\sigma_8$ and of $\Omega_m$, and they extend the fits to extremely small scales (80$h^{-1}$ kpc). These fits, however, do not explicitly express the distribution parameters in terms of cosmology.

\citet{Uhlemann2016} employ a first-principles approach to this problem; their results (from applying the Large Deviation Principle to spherical collapse) provide a good description of the density field in the low-variance limit. They find it helpful to consider the log density -- which we denote $A = \ln(1+\delta)$ -- rather than the overdensity $\delta$ itself. \citet{Repp2017} also consider the log field, but they employ a phenomenological approach (using the Millennium Simulation) to predict the power spectrum $P_A(k)$; they find that $P_A(k)$ essentially equals the linear power spectrum $P_\mathrm{lin}(k)$, biased by the ratio of the linear variance $\sigma^2_\mathrm{lin}$ to the $A$-variance $\sigma_A^2$.

Since $\delta$ is approximately lognormal, it is reasonable to suppose in extracting the cosmological information) that the first three moments of $A$ 
characterize the log density. \citet{Repp2017} have already characterized the variance of $A$; in this work we provide fits for its mean $\langle A \rangle$ and skewness $T_3$. These fits allow prediction of these moments for any (near-concordance) set of cosmological parameter values, including combinations of parameters that were not in the original fit.

We also show that the Generalized Extreme Value (GEV) distribution specified by these moments describes well the actual distribution of $A$. In particular, the GEV distribution determined by the predicted moment values is accurate to within 1.7 per cent (in near-concordance cosmologies) for redshifts up to $z \sim 2$ and for smoothing scales (pixel side lengths) from 2 to $30 h^{-1}\mathrm{Mpc}$.

We structure this paper as follows: in Section~\ref{sec:var} we reiterate our prescription for the variance of the log field and clarify the passage from $\sigma_A^2(k)$ to $\sigma_A^2(\ell)$. In Section~\ref{sec:mean} we present a similar prescription for the mean $\langle A \rangle$; and in Section~\ref{sec:skew} we present our prescription for the skewness $T_3$ of the log density field. In Section~\ref{sec:PDF} we present the GEV model for the $A$-probability distribution and quantify its accuracy. Discussion, including comparison to \citet{Uhlemann2016}, follows in Section~\ref{sec:disc};  we summarize and conclude in Section~\ref{sec:concl}.

\section{Variance}
\label{sec:var}
The over/under\-density of a location in the Universe is $\delta = \rho / \overline{\rho} - 1$,
where $\rho$ denotes density. We here consider the log density field
\begin{equation}
A = \ln(1+\delta).
\end{equation}
Since $\delta$ exhibits an approximately lognormal distribution, $A$ is Gaussian to first approximation.

Furthermore, the log transformation effectively erases non-linear evolution from the power spectrum \citep{NSS09} -- despite $A$ being only approximately Gaussian. Based on this surprising empirical fact, one can write the following relationship between the power spectrum of $A$ and the linear power spectrum predicted, for instance, by CAMB (Code for Anisotropies in the Microwave Background:\footnote{http://camb.info/} \citealp{CAMB}):
\begin{equation}
P_A(k) = b_A^2 P_\mathrm{lin}(k).
\label{eq:pwrspec}
\end{equation}
\citet{Repp2017} show that this relation holds to better than 10 per cent; the addition of a small slope modulation parameter reduces the error to a few per cent, comparable
to that of the standard \citet{Smith_et_al} fit.

It follows immediately that the bias $b_A^2$ equals the ratio between the variance of $A$ and the linear variance. Furthermore, since these variances occur in the context of Equation~\ref{eq:pwrspec}, they are functions of the wavenumber $k$ and are obtained by integrating the power spectra against a top-hat filter in $k$-space. We thus write these quantities in terms of the Nyquist wavenumber $k_N = \pi/\ell$, where $\ell$ is the side length of the cubical pixels; and we denote these quantities as $\sigma^2_A(k_N)$ and $\sigma^2_\mathrm{lin}(k_N)$, defined by
\begin{equation}
\sigma^2_{A,\mathrm{lin}}(k_N) \equiv \int_0^{k_N} \, \frac{dk\,k^2}{2\pi^2}P_{A,\mathrm{lin}}(k),
\label{eq:sigk}
\end{equation}
so that
\begin{equation}
b_A^2 = \frac{\sigma^2_A(k_N)}{\sigma^2_\mathrm{lin}(k_N)}.
\label{eq:bias}
\end{equation}

\citet{Repp2017} use snapshots\footnote{http://gavo.mpa-garching.mpg.de/Millennium/} from the Millennium Simulation \citep{Springel2005} to show that this $A$-variance is a simple function of linear variance:
\begin{equation}
\sigma_A^2(k_N) = \mu \ln \left(1 + \frac{\sigma_\mathrm{lin}^2(k_N)}{\mu} \right),
\label{eq:sigAfit}
\end{equation}
where the best-fitting value of $\mu$ is 0.73.

We stress that $\sigma_A^2(k_N)$ is \emph{not} the variance one would compute from counts-in-cells in a cosmological simulation. To elucidate this point, one can consider the $A$-field of the Millennium Simulation at $z=0$. Applying Equation~\ref{eq:sigk} to the measured power spectrum $P_A(k)$ of this field yields a variance of 0.966; the actual counts-in-cells variance is 1.49.

Two factors are responsible for this discrepancy. First, the counts-in-cells procedure introduces convolution with a cubical top-hat filter (mass-assignment function) in real space, which would require a factor of $W(k)^2$ in Equation~\ref{eq:sigk}. Second, and more subtly, this convolution occurs at only a finite number of points and is thus liable to alias effects from wavenumbers that are even-integer multiples of the Nyquist frequency. Both effects -- the convolution with the mass-assignment  function and the finite sampling of the convolved density field -- are significant near the Nyquist wavenumber.

To account for these effects, we follow \citet{Jing2005} in writing
\begin{equation}
P_{A\,\mathrm{meas}}(\mathbf{k}) = \sum_{\mathbf{n} \in \mathbb{Z}^3} P_A(\mathbf{k}+2k_N\mathbf{n}) W(\mathbf{k}+2k_N\mathbf{n})^2,
\label{eq:P_waa}
\end{equation}
where the sum runs over all three-dimensional integer vectors $\mathbf{n}$, and the Nyquist wavenumber $k_N = \pi/\ell$, $\ell$ being the distance between neighbouring grid points. (Note that we find it sufficient to consider only $|\mathbf{n}| < 3$. Note also that this equation is valid only for $|\mathbf{k}| \le k_N$.)

We now use $\sigma^2_A(\ell)$ to denote the measured counts-in-cells variance -- the result of applying a cubical top-hat filter in real space. Then
\begin{equation}
\sigma^2_A(\ell) = \int_{V_k\setminus\{0\}} \frac{d^3k}{(2\pi)^3} P_{A\,\mathrm{meas}}(\mathbf{k}).
\label{eq:sigR}
\end{equation}

Note that the region of integration (denoted $V_k\setminus\!\{0\}$) is the set of $k$-vectors corresponding to the real-space volume under consideration; since the mean of $A$ does not vanish, we must exclude the zero-wavenumber power $P_A(0)$. (Technically one should also exclude modes larger than the real-space length scale; in practice the impact of doing so is negligible.)

In particular, if the spatial volume is cubical (as in the Millennium Simulation), then $V_k$ is a cube in $k$-space extending from $-k_N$ to $k_N$ along each axis. In such a case, integration over a sphere (i.e., replacing $d^3k$ with $4\pi\,k^2\,dk$ and integrating up to $k_N$) yields a value for $\sigma^2_A(\ell)$ that is too small; obtaining the correct value requires integration over the entire cube.

One additional subtlety enters here, in that $V_k$ includes some vectors with magnitude greater than $k_N$, for which Equation~\ref{eq:P_waa} does not hold. Comparison with direct measurements from the simulation shows that a power law continuation of Equation~\ref{eq:P_waa} yields sufficiently accurate values of $\sigma_A^2(\ell)$.

\begin{figure}
    \leavevmode
    \includegraphics[width=9.2cm]{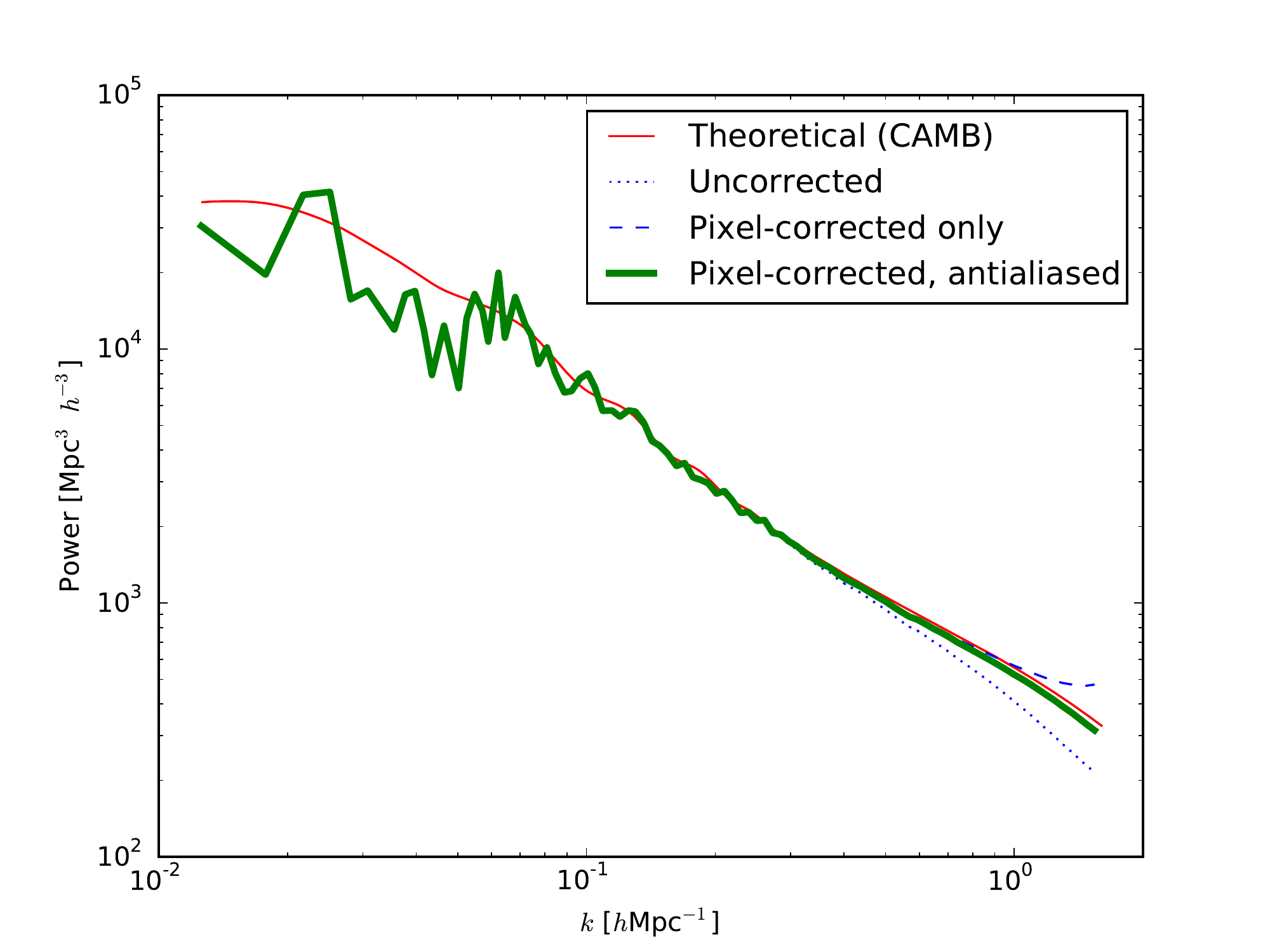}
    \caption{Power spectrum of the Millennium Simulation at $z = 0$, compared to the cosmic-mean prediction from CAMB. The dotted line shows the measured (uncorrected) power spectrum; the dashed line shows the measured spectrum corrected for the mass-assignment (pixel window) function only. It is clear that both pixel-window correction and antialiasing are necessary in order to recover the theoretical prediction \protect\citep{Jing2005}.}
\label{fig:spectra}
\end{figure}

\citet{Jing2005} also provides an iterative algorithm to de-alias and deconvolve the measured power, thus reconstructing the original spectrum. Fig.~\ref{fig:spectra} shows that this procedure is necessary to match the measured power spectrum $P(k)$ to that generated by CAMB. It is somewhat surprising that this procedure works for the $A$-power spectrum as well, since measuring $P_A(k)$ involves first smoothing and then taking the log, whereas Equation~\ref{eq:pwrspec} implicitly assumes that we first take the log -- to get the `true' $P_A(k)$ -- and then smooth it (to the measured pixel scale). These operations do not in general commute; however, fig. 2 of \citet{Repp2017} shows that \citeauthor{Jing2005}'s prescription succeeds in recovering the theoretical curve, whether one applies it to the measured spectrum of $\delta$ or of $A$.

Thus, returning to the variance measurements quoted following Equation~\ref{eq:sigAfit}, if we calculate the variance by integrating the measured $A$-spectrum (i.e., by using $P_{A\,\mathrm{meas}}$ instead of $P_A(k)$ in Equation~\ref{eq:sigk}), we obtain 0.966. If we use the deconvolved and antialiased $P_A(k)$, Equation~\ref{eq:sigk} gives us $\sigma_A^2(k_N) = 1.18$. However, the measured counts-in-cells variance $\sigma_A^2(\ell)=1.49$, and this is the variance which Equation~\ref{eq:sigR} recovers by integrating over the cube in $k$-space. The value of $\ell = \pi/k_N$ is the side length of the cubical pixels, which in this case is $1.95h^{-1}$ Mpc.

It is this variance $\sigma^2_A(\ell)$ which partially determines the probability distribution in Section~\ref{sec:PDF}. To predict it, one first obtains $\sigma^2_A(k_N)$ from Equation~\ref{eq:sigAfit}, noting that $k_N$ depends on the side length of a cubical pixel in real space. One then obtains $P_A(k)$ from Equations~\ref{eq:pwrspec}--\ref{eq:bias}, and then $\sigma^2_A(\ell)$ from Equations~\ref{eq:P_waa} and \ref{eq:sigR}.

\section{Mean}
\label{sec:mean}

Since a simple relation connects $\sigma_A^2(k_N)$ and $\sigma_\mathrm{lin}(k_N)$, it is reasonable to hope that a similar relation might connect $\langle A \rangle$ and $\sigma_\mathrm{lin}^2(k_N)$.

\begin{figure}
    \leavevmode
    \includegraphics[width=9.2cm]{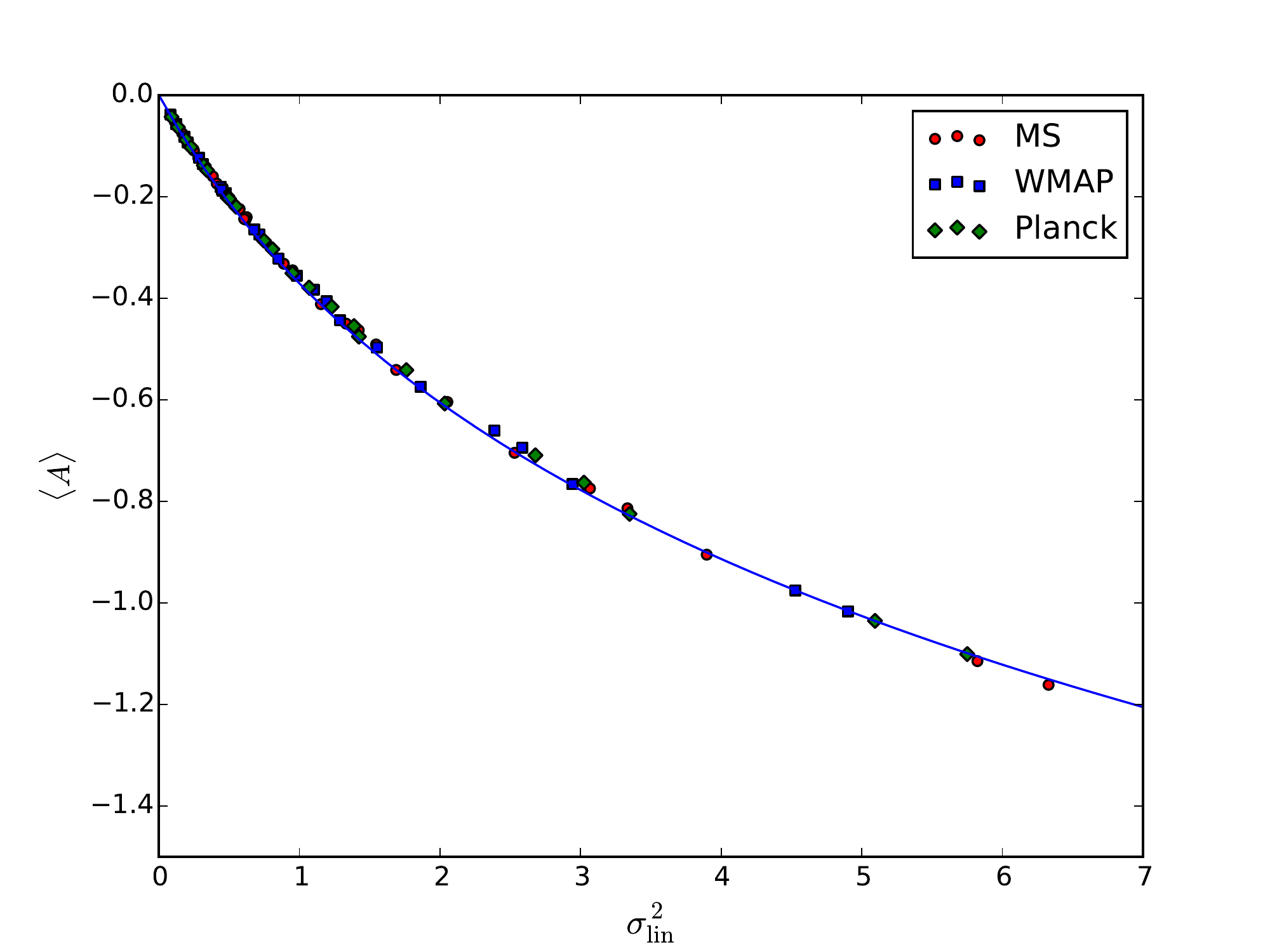}
    \caption{The mean of the log density field as a function of linear variance. The data points show values of $\langle A \rangle$ measured from the Millennium Simulation at redshifts ranging from $z = 0$ to 2.1 and smoothing scales from 2 to $32h^{-1}$ Mpc, in three different cosmologies (Planck 2013, WMAP7, and the original Millennium Simulation cosmology). The curve shows the best fit to Equation~\ref{eq:meanAfit} ($\lambda=0.65$).}
\label{fig:meanA}
\end{figure}

We begin by assuming that $\delta$ is lognormal in the low-variance limit, so that $\langle A \rangle = (-1/2)\sigma_A^2 \sim (-1/2)\sigma^2_\mathrm{lin}.$ Hence we attempt to fit the mean value of $A$ using the function
\begin{equation}
\langle A \rangle = -\lambda \ln \left(1+\frac{\sigma^2_\mathrm{lin}(k_N)}{2\lambda}\right);
\label{eq:meanAfit}
\end{equation}
once again, $k_N$ is the Nyquist wavenumber associated with the side length of the cubical pixels.

We find this form to be indeed a good fit for the mean of $A$ (see Fig.~\ref{fig:meanA}); least squares optimization on points with $\sigma^2_\mathrm{lin} > 2$ (to insure accuracy of the fit at high variances) yields a best value of $\lambda = 0.65$.

\begin{figure}
    \leavevmode
    \includegraphics[width=9.2cm]{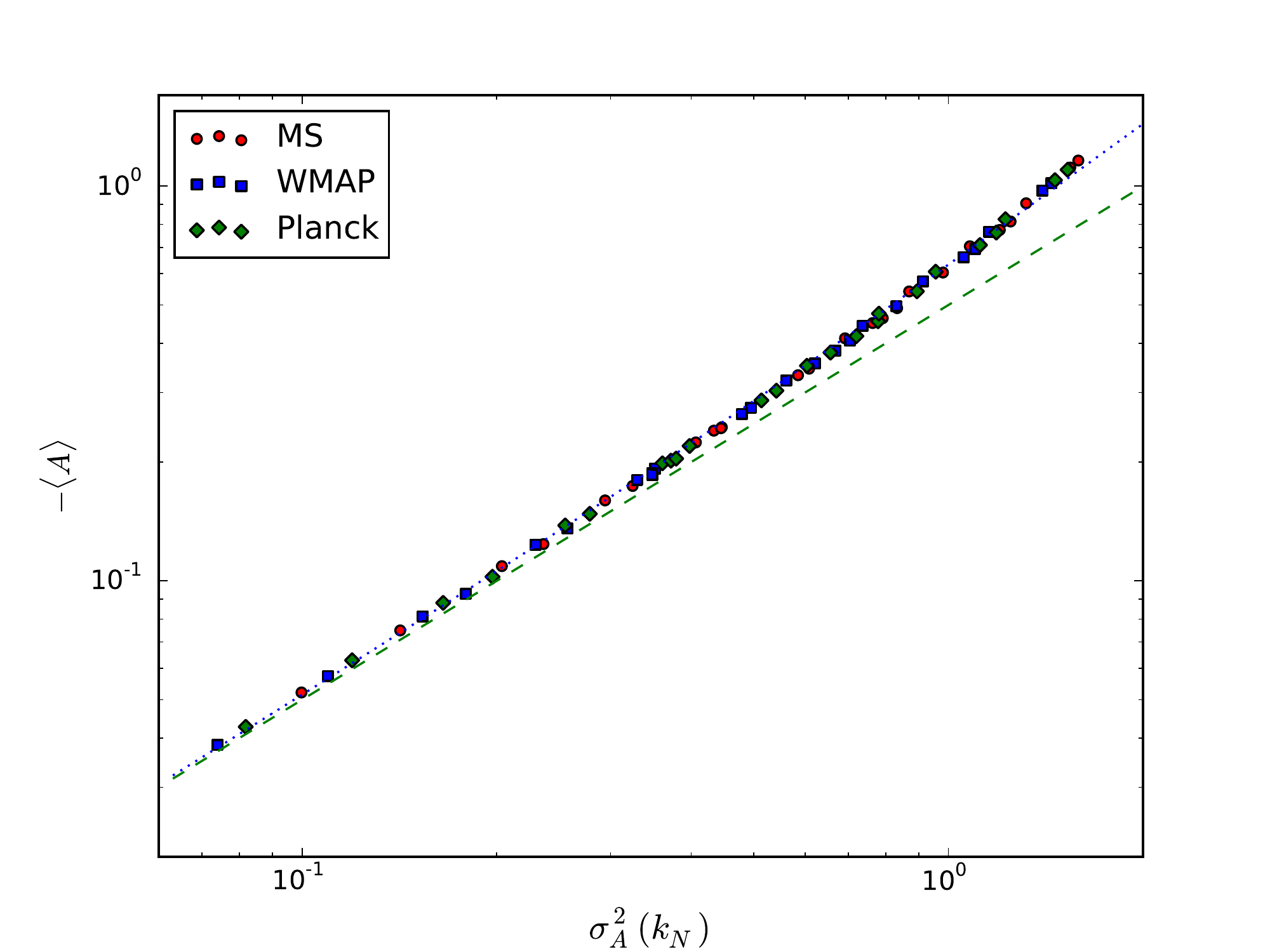}
    \caption{The mean of the log density field versus the variance of the same. The dashed line shows the lognormal prediction, which is a reasonable fit for low variances. The dotted curve shows the relationship implied by Equations~\ref{eq:sigAfit} and \ref{eq:meanAfit}. Data points are as in Fig.~\ref{fig:meanA}.}
\label{fig:meanAcomp}
\end{figure}
By combining Equations~\ref{eq:sigAfit} and \ref{eq:meanAfit}, we can express $\langle A \rangle$ directly as a function of $\sigma^2_A(k_N)$. When we do so (Fig.~\ref{fig:meanAcomp}), we see that in the low-variance regime ($\sigma_A^2(k_N) \la 0.1$) the mean is (by design) what one would expect from the lognormal approximation. At higher variances we see distinct departure from lognormal behavior.

\section{Skewness}
\label{sec:skew}
\begin{figure}
    \leavevmode
    \includegraphics[width=9.2cm]{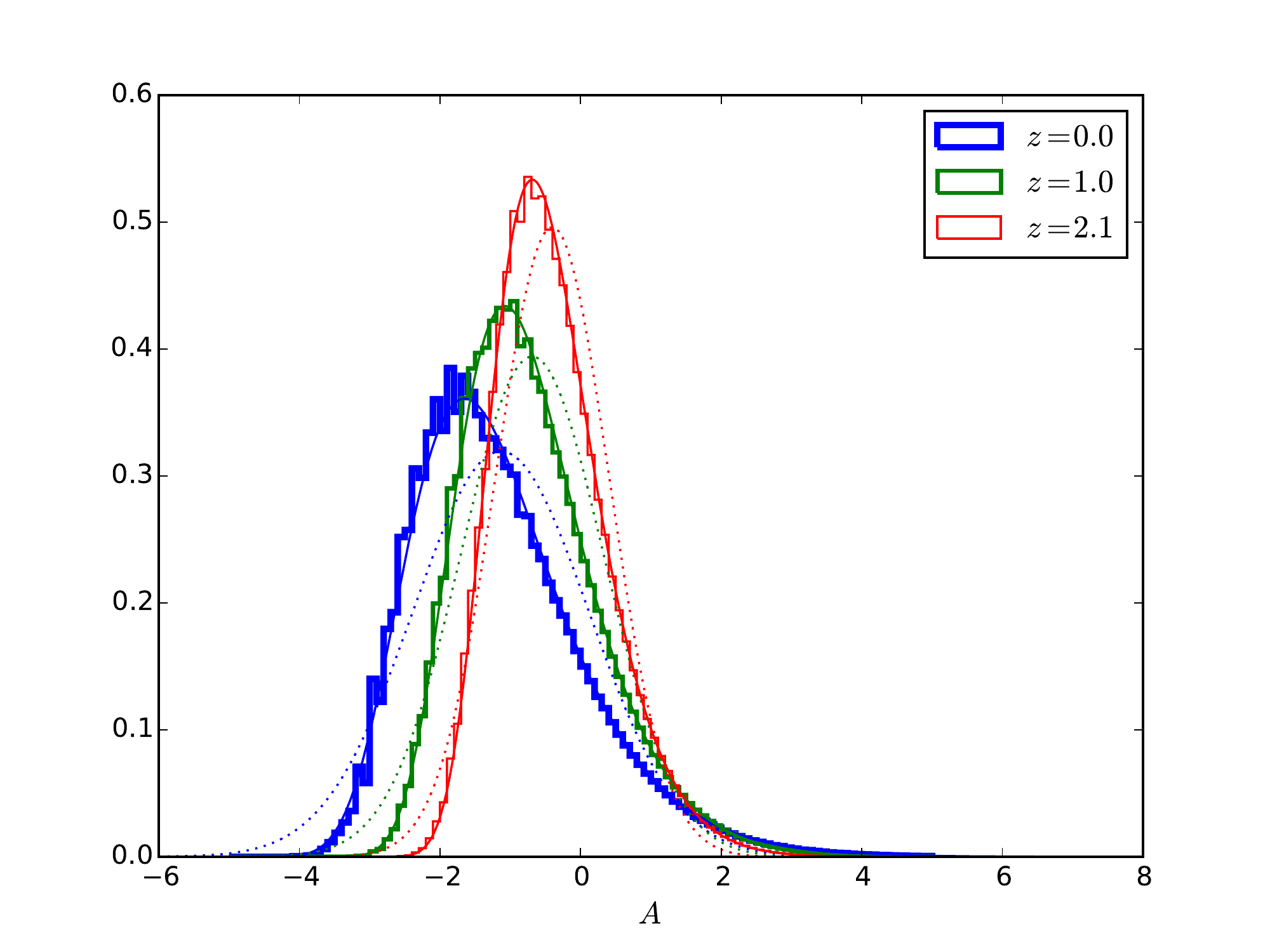}
    \caption{Histograms showing the log density probability distribution (from the Millennium Simulation) at various redshifts. The dotted curves show the best Gaussian fits with unconstrained mean. Skewness is apparent even at $z \sim 2$ and increases noticeably at later times, making the GEV fit (solid curves) a much better model for the log density. The pixel size in each case is $1.95h^{-1}$ Mpc.}
\label{fig:A_dist}
\end{figure}

For a truly lognormal distribution, the mean and variance of $A$ would be its only nonzero moments. However, it is well-known (e.g., \citealp{Colombi1994}) that the actual distribution of $A$ is noticeably skewed (see Fig.~\ref{fig:A_dist}). We thus turn now to characterizing this skewness.

The standard measure for the skewness of the density field is $S_3$, the third moment of $\delta$ scaled by the square of the variance:
\begin{equation}
S_3=\frac{\langle\delta^3\rangle}{\sigma^4}.
\end{equation}
We (following \citealp{Colombi1994}) denote the analogous measure for the log field as $T_3$:
\begin{equation}
T_3\equiv \frac{\langle\left(A-\langle A \rangle\right)^3\rangle}{\sigma_A^4(\ell)}.
\end{equation}
A related measure of skewness is the Pearson's moment coefficient $\gamma_1$:
\begin{equation}
\gamma_1 \equiv \frac{\langle\left(A-\langle A \rangle\right)^3\rangle}{\sigma_A^3(\ell)} = T_3 \cdot \sigma_A(\ell).
\end{equation}

\begin{figure*}
    \leavevmode
    \includegraphics[width=18cm]{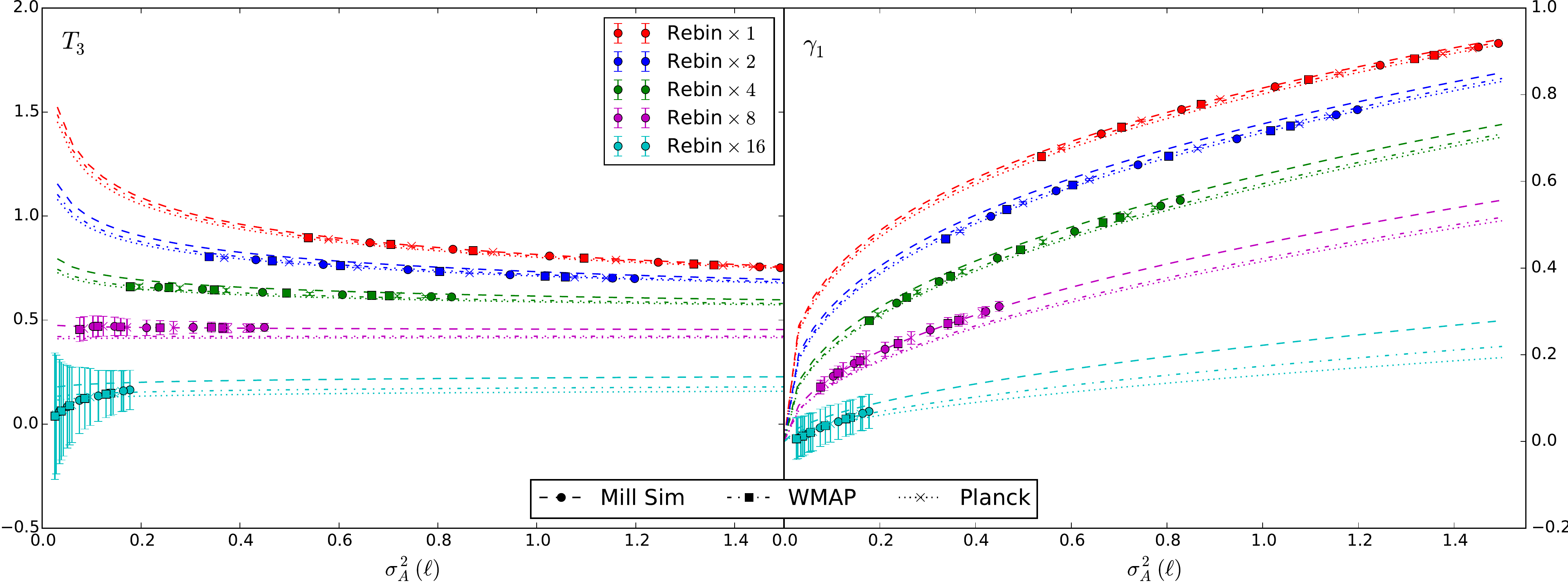}
    \caption{Left panel: values of $T_3$ measured in the Millennium Simulation for $z = 0.0, 0.1, 0.5, 1.0, 1.5$, and 2.1. Colours indicate rebinning factors (i.e., increases in pixel side lengths), and the three data point symbols indicate the three cosmologies (original Millennium Simulation, WMAP7, Planck 2013). We obtain the error bars by calculating $T_3$ in eight separate subcubes of the simulation. The curves show the predictions of our model; the type of curve (dashed, dash-dot, and dotted) distinguishes the cosmologies. As noted in the text, our fit exhibits a slight cosmology dependence which the simulation rescalings do not reflect. Right panel: values of $\gamma_1$ measured in the Millennium Simulation -- along with our fits -- for the same redshifts, smoothing scales, and cosmologies.}
\label{fig:T3_fits}
\end{figure*}

When we plot values of $T_3$ from the Millennium Simulation (see Fig.~\ref{fig:T3_fits}), we note that it is roughly constant at each scale; however, both our results and the simulations of \citet{Uhlemann2016} show a mild dependence of $T_3$ on the variance. We also consider that, according to a standard result from perturbation theory, $S_3 = 34/7 - (n_s +3)$, where $n_s$ is the slope of the linear power spectrum at the scale being considered. While one cannot carry over the derivation of this result into log space, it is reasonable to expect a similar dependence for $T_3$.

We thus propose an expression for $T_3$ that depends both on $\sigma_A^2(\ell)$ (which varies with redshift) and on the linear power spectrum slope $n_s$ at the scale $\ell$. By rebinning the Millennium Simulation into pixels with side lengths 2, 4, 8, and 16 times the original pixel size (and then calculating $A$ for the rebinned volumes), we obtain $A$-fields at a variety of redshifts (from 0 to 2.1) and multiple smoothing scales. These results are available in three ``near-concordance cosmologies,'' (viz., the original Millennium Simulation cosmology, the WMAP7 cosmology, and the Planck 2013 cosmology), using the publicly-available rescalings from the \citet{AnguloWhite} algorithm.

We now plot both $T_3$ and $\gamma_1$ as functions of $\sigma^2_A$ for the various redshifts, scales, and cosmologies described above; the results appear in Fig.~\ref{fig:T3_fits}.
As noted before, $T_3$ is roughly constant at each smoothing scale; however, both its magnitude and its dependence on $\sigma_A^2$ are sensitive to this scale. It is at first puzzling that (for small scales) $T_3$ appears to \emph{rise} at low variances, but this fact simply indicates that the third moment of $A$ approaches zero slightly more slowly that $\sigma_A^4(\ell)$; if we reduce by one the power of $\sigma_A(\ell)$ in the denominator, we obtain $\gamma_1$, which approaches zero as a power-law.

We thus write $T_3$ as a power law in $\sigma_A^2(\ell)$ with a coefficient $T(n_s)$; since $T_3$ is roughly constant at each scale, we expect the power $-p(n_s)$ to be relatively small. Thus,
\begin{equation}
T_3 = T(n_s)\cdot \left( \sigma_A^2(\ell) \right) ^ {-p(n_s)},
\label{eq:T3}
\end{equation}
where $n_s$ is the slope of the linear power spectrum. Note that we find it necessary to use the no-wiggle linear power spectrum of \citet{EisensteinHu}, since the inclusion of baryonic oscillations significantly affects the scale-dependence of the slope.
Roughly speaking, $T(n_s)$ is the constant part of the skewness, and the power expresses evolution with redshift.

It remains to write expressions for $T(n_s)$ and $p(n_s)$. We find that $T(n_s)$ depends linearly upon $n_s$, and that $p(n_s)$ is well-fit by a logarithmic function:
\begin{eqnarray}
T(n_s) & = & a(n_s+3) + b \label{eq:Tns}\\
p(n_s) & = & d + c \ln(n_s+3).
\end{eqnarray}
Note that $b$ is the value in the limit of $n_s=-3$, corresponding to $34/7$ in the standard result for $S_3$.

We then perform MCMC optimization (over the measured values of $T_3$ for our six redshifts, five smoothing scales, and three near-concordance cosmologies) to obtain the following best fit values of $a$, $b$, $c$, and $d$:
\begin{equation}
\label{eq:T3params}
\begin{array}{ccccccc}
a & = & -0.70 & \hspace{1cm} & c & = & -0.26\\
b & = & 1.25  & \hspace{1cm} & d & = & 0.06.
\end{array}
\end{equation}
Note that relative to the expression for $S_3$, the coefficient of $n_s+3$ has changed from $-1$ to $-0.7$, and the constant term has decreased from $34/7$ to 1.25 -- indicating the reduction in skewness accomplished by the log transformation.

We plot this fit (for each of our three cosmologies) in Fig.~\ref{fig:T3_fits}. While noting that it captures the trend of the data, there is scope for improvement in two areas. First, the algorithm of \citet{AnguloWhite} -- by which the WMAP and Planck results were derived from the original Millennium Simulation -- rescales the simulation volume and reassigns the redshifts of each snapshot to match the variance in a range of redshifts and scales. This procedure reproduces the power spectrum of the target cosmology quite accurately. It is also responsible for the fact that at each rebinning factor, the data points from all three cosmologies line up in Fig.~\ref{fig:T3_fits}: for instance, $z = 0$ corresponds to a different snapshot in each of the three cosmologies, but these three snapshots come from the \emph{same} evolutionary sequence in the same simulation. However, the closely parallel (but not coincident) curves in Fig.~\ref{fig:T3_fits} show that our use of $n_s$ to parametrize the scale-dependence introduces a small cosmology-dependence that differs slightly from \citeauthor{AnguloWhite}'s rescaling. Understanding (and, if necessary, correcting) this difference requires further testing in other simulations (see Section~\ref{sec:disc}).
 
Second, the low-variance data in Fig.~\ref{fig:T3_fits} suggest a possible downturn in $T_3$ as $\sigma_A^2$ approaches zero. We considered including a factor in Equation~\ref{eq:T3} to model this downturn; however, doing so would introduce additional free parameters only in order to model a regime dominated by large error bars. Thus we leave the existence of and modeling of any such downturn for future work, given that Equation~\ref{eq:T3} is sufficiently accurate for predicting the distribution of $A$.

However, it is worth noting once again how dramatically the log transform reduces the effects of non-linear evolution. The perturbative expression for $S_3$, viz. $34/7-(n_s+3)$, fails catastrophically at the scale of the full-resolution Millennium Simulation: at this scale, the measured value of $S_3$ (at $z = 0$ in the original cosmology) is $8.4 \pm 0.3$, twice as large as the value of 4.2 predicted by the perturbative expression. At the largest scale we consider (cubical pixels of side $31.25h^{-1}$ Mpc), the perturbative prediction for $S_3$ is better, yielding 3.4 as opposed to the measured value of $2.8 \pm 0.2$. The analogous expression for $T_3$, with no explicit scale-dependence, is Equation~\ref{eq:Tns}. On our smallest scale it predicts a value of 0.81, compared to a measured value of $0.75$ (uncertainty $<.01$); on the largest scale, it predicts a value of 0.21, consistent with the measured value of $0.16 \pm .09$. In this regard the log-transformed distribution exhibits much more nearly linear behavior even down to scales of $2h^{-1}$ Mpc.

In addition, \citet{Neyrinck2013} notes that in a variety of distributions, including the lognormal, the log transform reduces the skewness by three (so that $T_3 = S_3 - 3$). In our fits we find behavior consistent with this relationship at the largest scale ($S_3 = 2.9 \pm 0.2$, $T_3 = .16 \pm .09$), though not at the smallest ($S_3 = 8.3 \pm 0.3$, $T_3 = 0.75$). Furthermore, \citet{Uhlemann2016} show that the disparity between $T_3$ and $S_3 - 3$ is proportional to $\sigma_A^2$ (plus terms of higher order in $\sigma_A^2$). This dependence on the $A$-variance explains the fact that the smallest scale ($\sigma_A^2 = 1.5$) exhibits significant departure from the relationship, whereas the largest scale ($\sigma_A^2 = 0.18$) does not.

\section{Probability Distribution}
\label{sec:PDF}
With expressions in hand for the mean, variance, and skewness of the log density field, it remains to model the actual probability distribution of $A$. We begin by noting that one does not directly sample the actual, continuous $A$ (or $\delta$) distribution; instead one first smooths the densities across each pixel before sampling. This smoothing is tantamount to summing the underlying density across the points within that pixel.

If these intrapixel densities were uncorrelated, then we would expect a Gaussian result by the Central Limit Theorem. In constrast, we know that densities are well-correlated on scales much lower than our pixel size, and therefore one would expect a different limiting distribution. \citet{BertinClusel} demonstrate that extreme value statistics can describe sums of correlated variables, even if the underlying process is not extremal.

There are three classes of such extreme-value distributions, namely, the Gumbel, Fr\'{e}chet, and reversed Wiebull distributions (see, e.g., \citealp{GEVGumbel, GEVColes, GEVLB}). Of these three distributions, the Gumbel exhibits the simplest form and possesses the advantage of support at all real numbers. \citet{Antal2009} note that the galaxy counts from SDSS seem to follow a Gumbel distribution, and this distribution is also a fairly good fit to the Millennium Simulation distribution of $A$ at low redshifts (without any rebinning).

However, the skewness ($\gamma_1$) of the Gumbel distribution is fixed at $\sim 1.1$, and thus it is not a good fit for $A$ at higher redshifts or at different smoothing scales. We obtain a better fit with the reversed Wiebull class of Generalized Extreme Value (GEV) distributions, which allows for variable skewness. The GEV distributions depend on three parameters -- a location parameter $\mu_\mathrm{GEV}$, a positive scale parameter $\sigma_\mathrm{GEV}$, and a shape parameter $\xi$. The probability distribution is then
\begin{equation}
\label{eq:GEV}
\mathcal{P}(A) = \frac{1}{\sigma_\mathrm{GEV}} t(A)^{1+\xi} e^{-t(A)},
\end{equation}
where
\begin{equation}
\label{eq:GEV_t}
t(A) = \left(1 + \frac{A - \mu_\mathrm{GEV}}{\sigma_\mathrm{GEV}}\xi\right)^{-1/\xi}.
\end{equation}
The Gumbel distribution is the limit of this expression as $\xi$ approaches zero, and it is clear that a double exponential is the most salient feature of the distribution in this limit.

The reversed Wiebull subclass of the GEV denotes distributions for which the shape parameter $\xi < 0$. In this case, the distribution parameters depend on the moments of $A$ as follows:
\begin{equation}
\gamma_1 = - \frac{\Gamma(1-3\xi)-3\Gamma(1-\xi)\Gamma(1-2\xi)+2\Gamma^3(1-\xi)}{\left(\Gamma(1-2\xi)-\Gamma^2(1-\xi)\right)^{3/2}};
\end{equation}
from this equation one can obtain $\xi$ numerically, and then the following two equations complete the specification of $\mathcal{P}(A)$:
\begin{equation}
\sigma_\mathrm{GEV} = \xi \sigma_A(\ell) \left( \Gamma(1-2\xi) - \Gamma^2(1-\xi)\right)^{-1/2}
\end{equation}
\begin{equation}
\mu_\mathrm{GEV} = \langle A \rangle - \sigma_\mathrm{GEV} \frac{\Gamma(1-\xi)-1}{\xi}.
\label{eq:GEV_mu}
\end{equation}

This distribution has support only for $A \le \mu_\mathrm{GEV} - \sigma_\mathrm{GEV}/\xi$; it is defined to be zero for any larger values of $A$. Thus this model implies an upper bound on a region's density at a given epoch and scale.

\begin{figure*}
    \leavevmode
    \includegraphics[width=18cm]{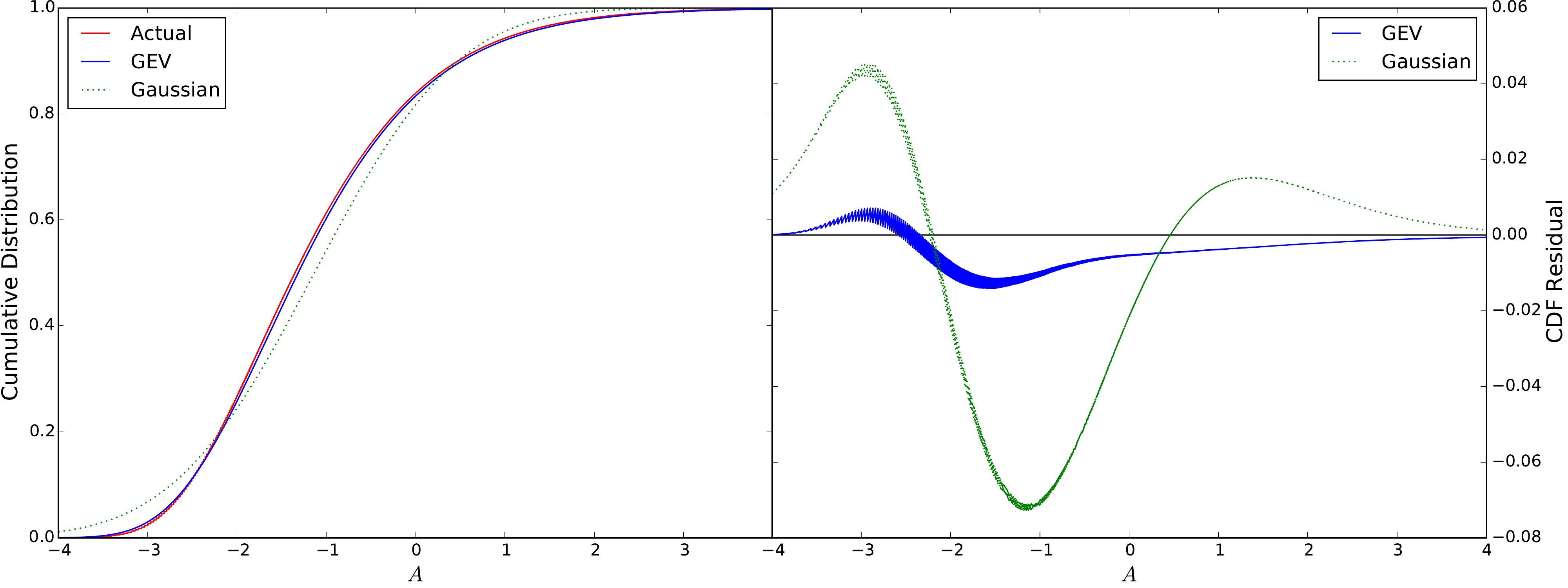}
    \caption{Left panel: cumulative distribution functions for the log density field $A$ (at $z=0$ in the original Millennium Simulation cosmology at a $1.95h^{-1}$ Mpc smoothing scale). The green curve shows the CDF for a Gaussian with the correct mean and variance, the blue curve shows the result of the General Extreme Value (GEV) prescription in this work, and the red shows the actual distribution produced by the simulation. Right panel: the residuals of the Gaussian and GEV prescriptions.}
\label{fig:CDF_compare}
\end{figure*}

This GEV distribution, together with our prescriptions for the mean, variance, and skewness of $A$, form a complete model for the log density distribution. We show the GEV fits (using moment values from Equations~\ref{eq:sigR}, \ref{eq:meanAfit}, and \ref{eq:T3}) in Fig.~\ref{fig:A_dist}. Inspection of this figure shows that the GEV distribution is a significantly better model than the Gaussian. And of course, given an accurate prescription for $A$, it is trivial to write the probability density for $\delta=e^A-1$.

To investigate the accuracy of this model, we compare its cumulative distribution function (CDF) to that of the Millennium Simulation realizations of $A$; we do so for our set of 6 redshifts, 5 smoothing scales, and 3 near-concordance cosmologies. Fig.~\ref{fig:CDF_compare} shows one such comparison; we note again the superiority of the GEV model to the Gaussian.

\begin{figure*}
    \leavevmode
    \includegraphics[width=18cm]{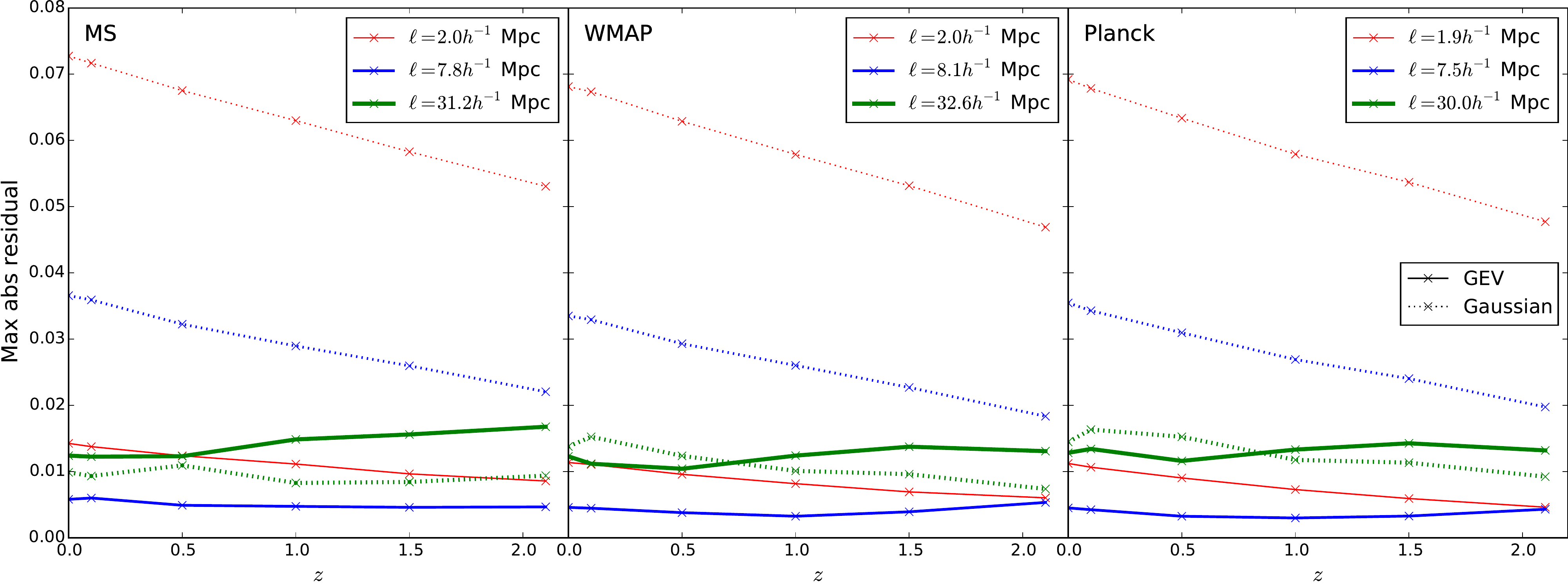}
    \caption{Solid lines -- maximum absolute differences between the cumulative distribution functions from the Millennium Simulation and from our GEV prescription; dotted lines -- maximum absolute differences between Millennium Simulation CDFs and those of Gaussian fits. The three panels show the three near-concordance cosmologies at a variety of redshifts and smoothing scales.}
\label{fig:KS_results}
\end{figure*}
We explore two means of quantifying the differences between the true and predicted CDFs. We first consider the maximum of the absolute difference between the CDFs; this maximum is in fact the Kolmogorov-Smirnov statistic. (See discussion of the K-S test in Section~\ref{sec:disc}.)
Fig.~\ref{fig:KS_results} shows these maximum residuals, and we note that the most extreme deviation is 1.7 per cent. This deviation compares quite favorably with the Gaussian distribution, whose worst deviation exceeds 7 per cent.

\begin{figure*}
    \leavevmode
    \includegraphics[width=18cm]{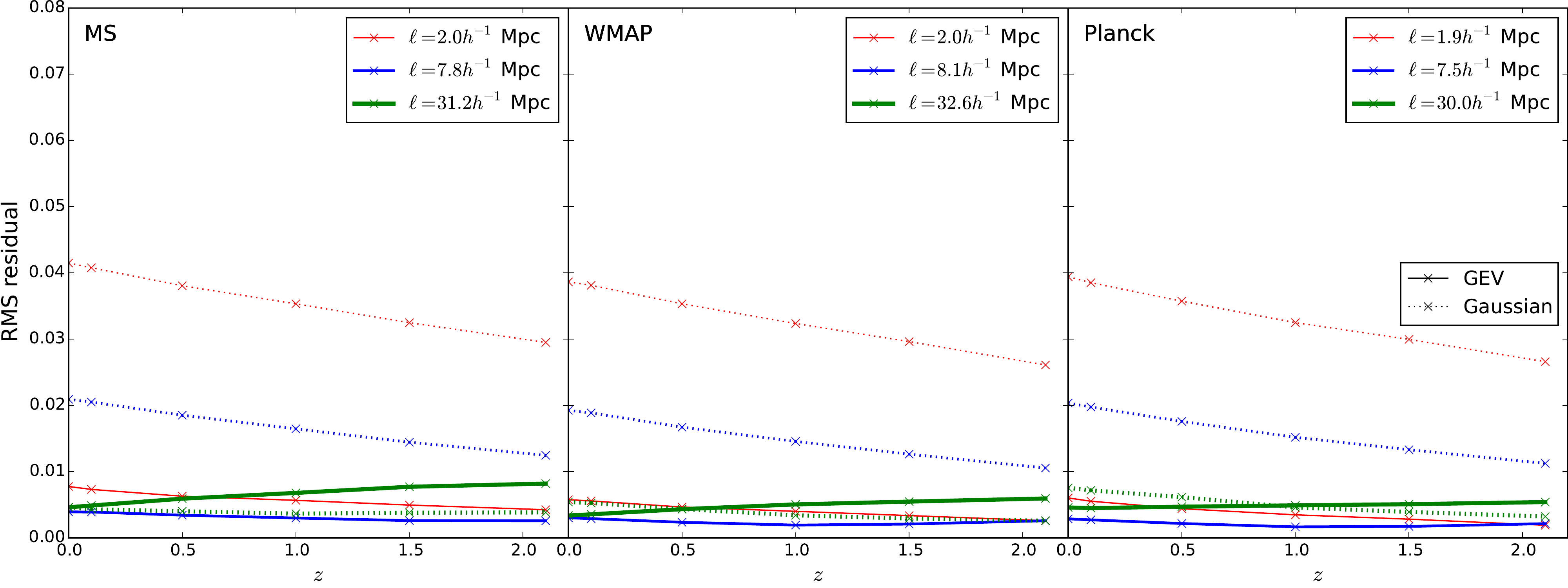}
    \caption{Root mean square differences between the cumulative distribution functions from the Millennium Simulation and those from GEV (and Gaussian) prescriptions. Panels and curves as in Fig.~\ref{fig:KS_results}.}
\label{fig:rms_results}
\end{figure*}
We next consider the root mean square (rms) values of the CDF differences (for the same set of cosmologies, redshifts, and smoothing scales). To prevent the vanishing residuals in the tails from dominating the rms, we consider only the values of $A$ for which the simulation CDFs fall between 0.05 and 0.95. The results appear in Fig.~\ref{fig:rms_results}; the worst GEV rms is 0.8 per cent, whereas the worst Gaussian rms is around 4 per cent. In this sense we can say that our prescription describes the log density distribution to sub-per cent levels of accuracy.

\section{Discussion}
\label{sec:disc}

In several ways our approach is complementary to that of \citet{Uhlemann2016}: they employ an a priori approach, starting from the assumption of spherical collapse and employing the Large Deviation Principle; our approach on the other hand is phenomenological.

In addition, their prescription applies to the low-variance limit, whereas ours is derived from $N$-body simulations on scales where the variance is of order unity. We have already noted (Section~\ref{sec:skew}) one indication that our prescription for $T_3$ might require correction in the low-variance regime. One might also deduce this fact from Figs.~\ref{fig:KS_results} and \ref{fig:rms_results}, which show that at large scales ($\sim 30h^{-1}$ Mpc) the GEV prescription in general does no better than the Gaussian. This scale ($k \sim 0.1$) represents the transition to the linear regime. It is this low-variance limit which \citet{Uhlemann2016} have well-characterized, and our prescription thus complements theirs by describing the mid- to high-variance regimes.

\begin{figure}
    \leavevmode
    \includegraphics[width=9.2cm]{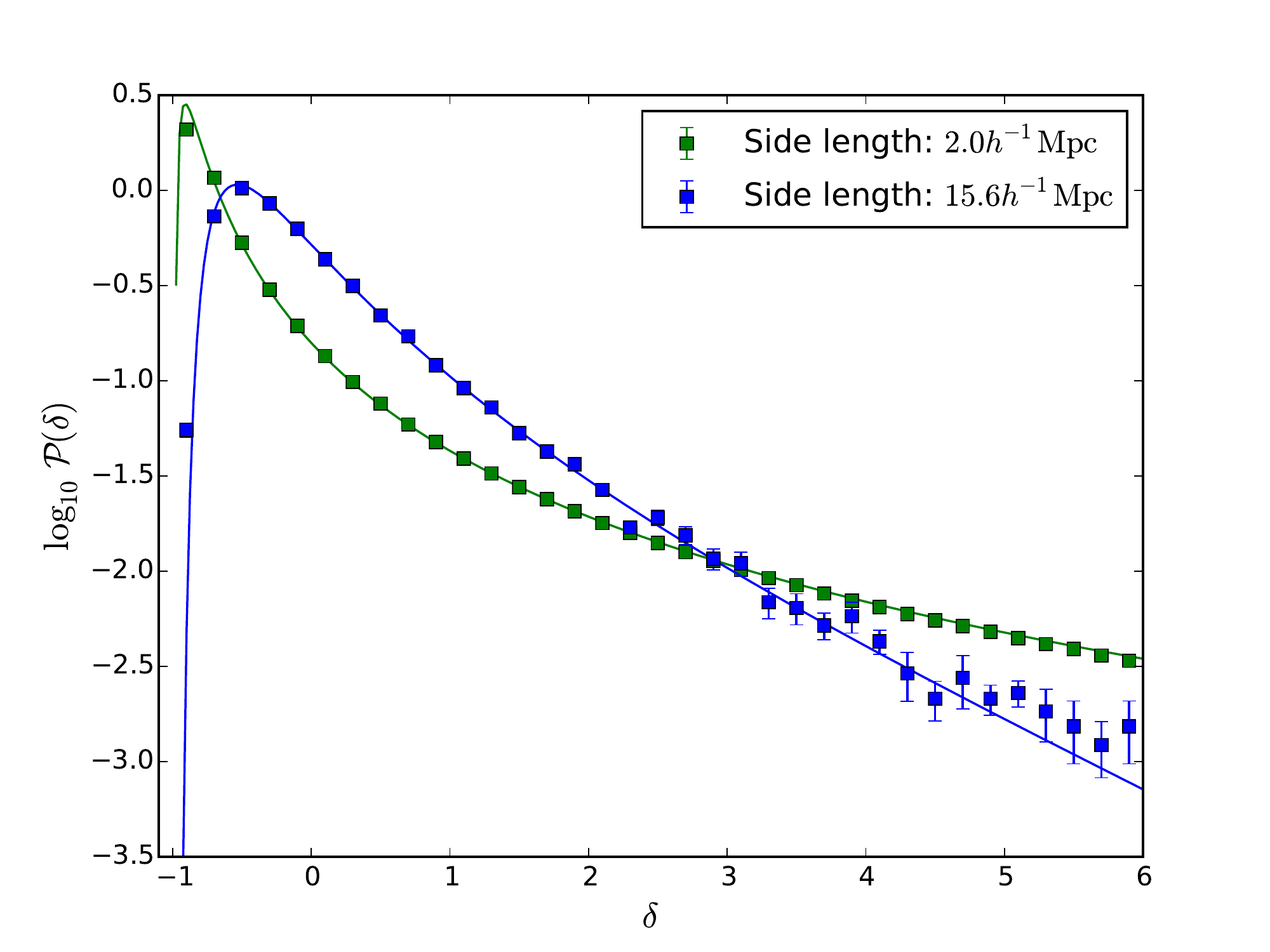}
    \caption{Predicted probability distribution functions of the overdensity field $\delta$ compared to the actual PDF measured from the Millennium Simulation (original cosmology) at $z = 0$. We show two different smoothing scales. We obtain error bars by calculating values in eight separate subcubes of the simulation volume.}
\label{fig:delta}
\end{figure}
However, the two results are similar in that both prescriptions predict double-exponential behavior for $\mathcal{P}(A)$ (see Appendix). And although the functional forms differ significantly, the results yield similar predictions for the distribution of $\delta$ (compare Fig.~\ref{fig:delta} with their fig.~8).

\begin{figure*}
    \leavevmode
    \includegraphics[width=18cm]{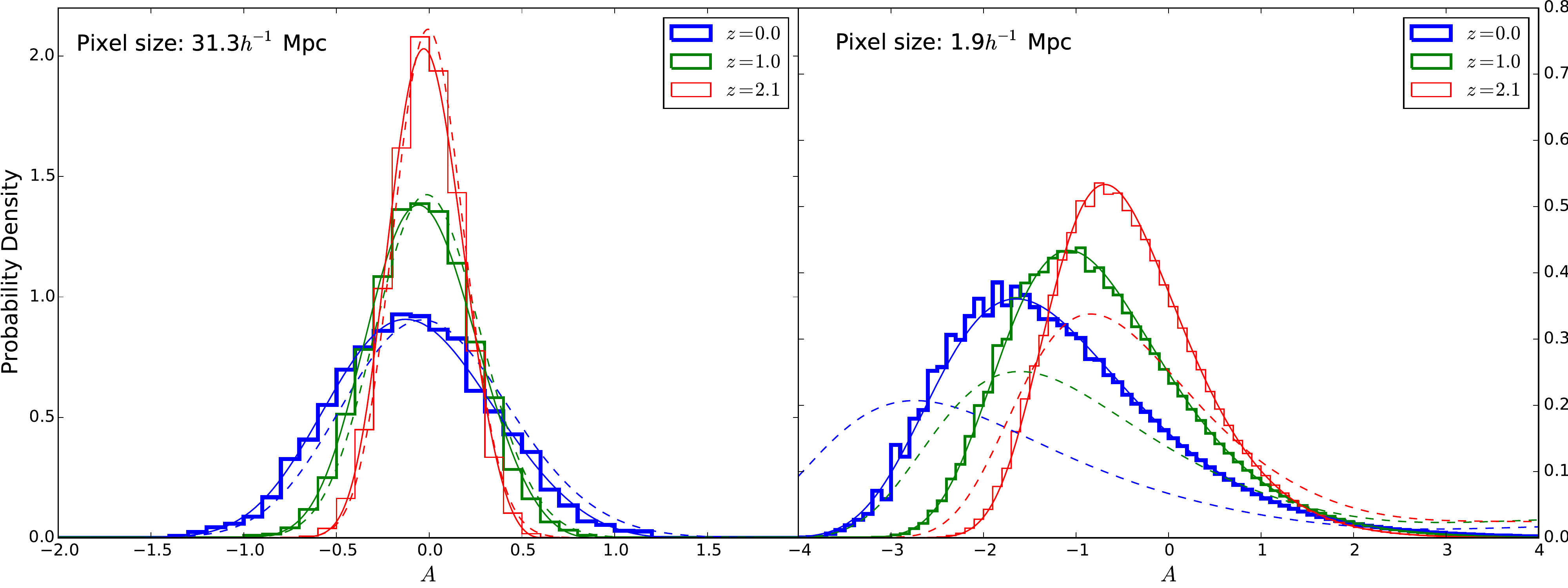}
    \caption{Predicted distributions of $A$ using our phenomenological fit (solid curves) and \protect\citet{Uhlemann2016}'s a priori approach (dashed curves); we compare both to distributions obtained from the Millennium Simulation at two smoothing scales. \protect\citeauthor{Uhlemann2016}'s prescription is applicable only in the low-variance limit (i.e., large scales and/or redshifts); in this limit (represented by the left-hand panel) the prescription presented in this work largely mimics their first-principle results.}
\label{fig:compareUCPBR}
\end{figure*}

In addition, our prescription approaches that of \citeauthor{Uhlemann2016} in the low-variance limit (see left-hand panel of Fig.~\ref{fig:compareUCPBR}, in particular the $z=2.1$ distribution). Thus, for situations in which their approach is valid, our GEV prescription reproduces the phenomenology predicted by their approach.

We next note three areas in which our fit is amenable to improvement. First, we have already observed (in Section~\ref{sec:skew}) the possibility of modifying our $T_3$-prediction at low variances.

Second, Fig.~\ref{fig:CDF_compare} shows that our model overpredicts the probability for small values of $A$ and underpredicts it for intermediate and large values. This effect is not peculiar to the redshift shown in Fig.~\ref{fig:CDF_compare}; one can observe it across the entire range of redshifts and smoothing scales. It is worth noting that when we have an extremely large number of simulation data points (as in the non-rebinned snapshots), the Kolmogorov-Smirnov test discriminates decisively between the GEV and the actual $A$ distribution. This result confirms that while the GEV is a good fit (with error less than 2 per cent), it is not a perfect fit.

Third, this prescription allows us to predict the fourth moment $T_4$ of $A$. Doing so, we find that the predicted values differ from the measured values by at most $\sim 0.2$. In many cases, this discrepancy is comparable to the size of the measurment error bars, and in the cases for which it is not, it represents an overprediction of about 20 per cent. We conclude that our prescription yields values for $T_4$ that are of the correct order of magnitude but cannot form the basis for a precise prediction.

Turning now to the underlying rationale for the GEV behavior of $A$, at least two possible scenarios could produce such a distribution. Both rely on the fact that our density measurements are averages over the points in a given pixel.

The first possibility (indicated by the name of the distribution) is a process that selects extreme values from the multiple realizations (at the various points in each pixel) of the underlying distribution. If the extremes of density dominate the log smoothed density in each cell, then one could expect GEV behavior. However, in that case one would also expect this behavior to be more pronounced at \emph{larger} smoothing scales, which integrate over a larger volume and thus include more-pronounced extremal values. In practice, we observe the opposite behavior: at large scales, $A$ exhibits a more nearly Gaussian distribution than at small scales. Furthermore, some simple investigations (into the effect of clipping extreme values before smoothing) indicate that the extreme values are not by themselves responsible for the GEV behavior of the log density distribution.

We have already mentioned (in Section~\ref{sec:PDF}) a second potential source of GEV behavior, namely, correlation of densities within a pixel. According to the Central Limit Theorem, the average of multiple realizations of independent identically distributed (and well-behaved) random variables approaches a Gaussian; however, correlations between the variables (violating independence) can yield extreme value statistics instead \citep{BertinClusel}. In this case, one would expect more-nearly Gaussian behavior on large scales as opposed to small scales, given that the correlations are weaker on larger scales. The Millennium Simulation pixel side lengths ($\sim 2h^{-1}$ Mpc) have the same order of magnitude as the typical correlation scale, and thus it is likely that the strong intrapixel correlations are the source of the GEV behavior in $A$. We leave further investigation of this possibility (using finer-grained simulations) to future work.

We next note that the parameters $\mu$ and $\lambda$ (in Equations~\ref{eq:sigAfit} and \ref{eq:meanAfit}) characterize non-perturbatively the deviations from lognormality. The variance is larger than that of a lognormal distribution, and the difference between $\mu$ and $\lambda$ corresponds to a distortion of the lognormal relationship between the mean and the variance. The Millennium Simulation (with its rescalings to two other near-concordance cosmologies) suggests that $\mu$ and $\lambda$ are universal, cosmology independent quantities characterizing the non-linear evolution. While this statement must be checked in wider range of cosmologies, it certainly appears to be true near the concordance model. These investigations we also leave to future research.

Finally, two distinct biases intervene between knowledge of the matter distribution and direct comparison to galaxy counts: the first reflects the impact of discreteness, and the second reflects the processes of galaxy formation and evolution.

\begin{figure*}
    \leavevmode
    \includegraphics[width=18cm]{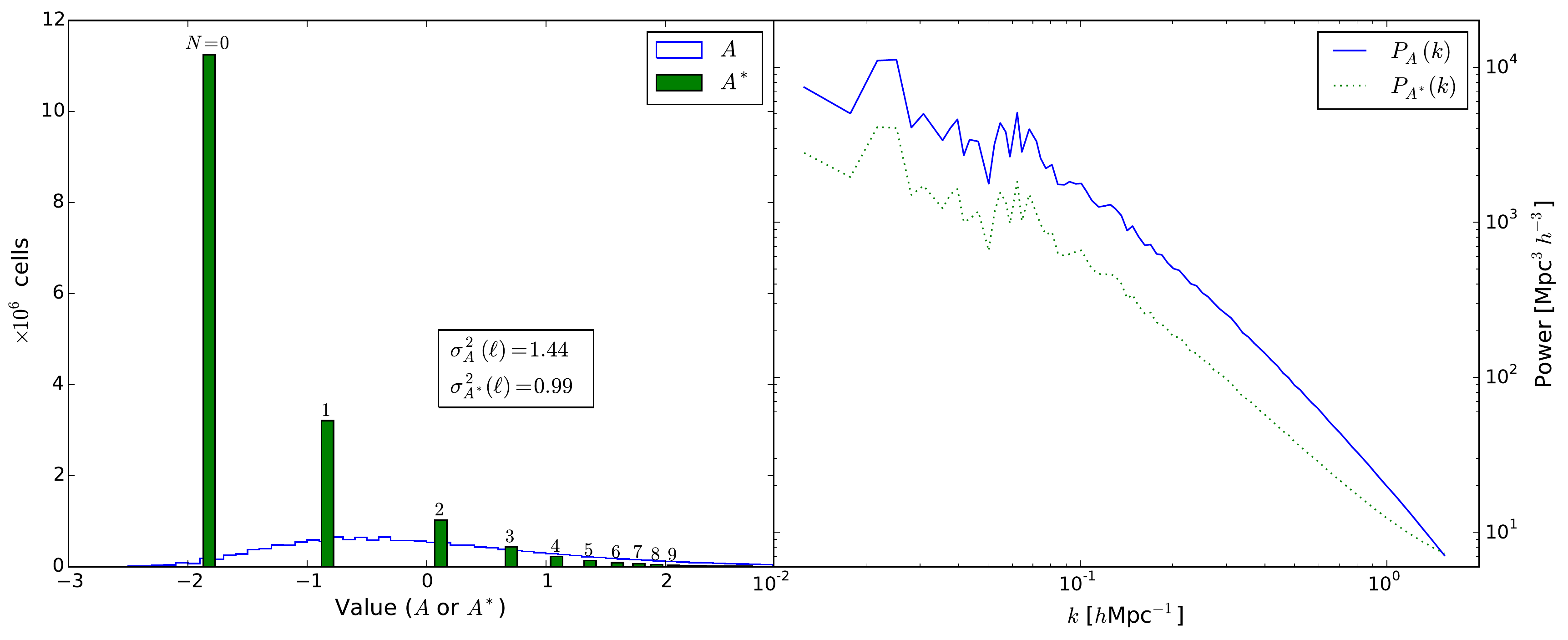}
    \caption{Left panel: Histogram of $A$- and $A^*$-distributions (both with bins of width 0.1). The $A$-distribution is measured from the Millennium Simulation (original cosmology) at $z=0$ in cubical cells of side length $1.95h^{-1}$ Mpc. The $A^*$-distribution reflects a random Poisson realization of the underlying $A$-field with mean number density $\overline{N}=1$ particle per pixel. The numbers above the first ten $A^*$-values indicate the corresponding galaxy number counts. Note that the passage from $A$ to $A^*$ reduces the variance by about 30 per cent. Right panel: Power spectra of the $A$- and  $A^*$-fields, showing both the $A^*$-bias and the high-$k$ discreteness plateau. Note that upon integration, the high-$k$ plateau ameliorates the effect of the bias so that the ratio of the variances is less pronounced than the ratio of the (low-$k$) power spectra.}
\label{fig:Astar}
\end{figure*}

Regarding the former, it is $A^*$, the discrete analog of $A$, which is the essentially optimal observable for discrete fields \citep{CarronSzapudi2014}. Since cosmological surveys count galaxies rather than directly measuring matter, it is the $A^*$ statistic that will ultimately reveal the information which escapes the standard power spectrum at high wavenumbers. However, the power spectrum of $A^*$ is biased with respect to that of $A$. Fig.~\ref{fig:Astar} shows the distribution and power spectrum of $A$ as measured in the Millennium Simulation at $z = 0$ in cubical pixels of side length $1.95h^{-1}$ Mpc. To generate a corresponding discrete realization, we first choose an average pixel number density of $\overline{N} = 1$ count per pixel; then for each pixel we randomly assign its number of counts $N$ from a Poisson distribution with mean $\overline{N}e^A = \overline{N}(1+\delta)$. After calculating $A^*(N)$ for each cell, we obtain the one-point distribution and power spectrum of $A^*$, both shown in the figure. In determining the power spectra, we use \citet{Jing2005}'s prescription to remove pixel-window and alias effects.

One notices first that $\sigma_{A^*}^2(\ell)$ is about 30 per cent lower than $\sigma_{A}^2(\ell)$, as expected given that much of the negative tail of the $A$-distribution collapses into the discrete $N=0$ spike. This reduced variance manifests itself in a power spectrum for $A^*$ that is biased with respect to that of $A$. One notices also that $P_{A^*}(k)$ begins to level off at high wavenumbers (an effect similar but not identical to the standard $1/\overline{n}$ discreteness plateau). Upon integrating the power spectra to obtain the variances, this upward bend partially ameliorates the effect of the bias; and thus the bias is even more pronounced than one might expect from the ratio of the variances. Approximate expressions for this bias exist \citep{WCS2015}, but they require further refinement. We have shown (Repp \& Szapudi, submitted) that the description of the $A$ field appearing in this work allows quantitative prediction of this bias $b_{A^*}^2$; and this bias is necessary to characterize the power spectrum of $A^*$. In addition, by predicting the probability distribution $\mathcal{P}(\delta)$, our prescription allows us to predict the distribution $\mathcal{P}(N)$ in simulations with lower particle density than the Millennium, and thus to further compare our results to those of other simulations.

This further comparison is important: thus far we have based all our results on the Millennium Simulation because its high particle density allows us to ignore discreteness effects. Unfortunately, the relatively small size of this simulation makes it susceptible to cosmic variance. Reference to other simulations will both alleviate this problem and expand the range of cosmologies which we can test. 

A second bias is the galaxy bias function defined by $\delta_g = f(\delta)$, where $\delta_g$ is the galaxy overdensity. \citet{SzapudiPan2004} have outlined and tested a method for determining this bias function from galaxy catalogs given the underlying dark matter distribution; one issue with which they contended was how to reconstruct (or what to assume for) that underlying distribution. Our prescription obviates this problem, allowing accurate characterization of the galaxy bias function; in particular, it permits such characterization apart from the assumption of a particular form (e.g., linear) for this bias.

\section{Conclusion}
\label{sec:concl}
The lognormal distribution is a reasonable first approximation to the matter distribution of the Universe. However, it is not an extremely accurate approximation, nor has there been until now a robust means of predicting the distribution parameters. This work remedies both problems, measuring and fitting the first three moments of the log distribution and showing that the GEV accurately describes $A = \ln(1+\delta)$.
Together with the GEV prescription, these fits allow the prediction of these moments -- and the one-point matter density distribution -- for any (near-concordance) set of cosmological parameter values. 

Both the mean $\langle A \rangle$ and the variance $\sigma_A^2(k)$ depend in a simple fashion on the linear variance $\sigma^2_\mathrm{lin}(k)$ (Equations \ref{eq:meanAfit} and \ref{eq:sigk}--\ref{eq:sigAfit}), and the dependence reduces to the expected lognormal behavior at low variances. However, the variance $\sigma_A^2(k)$ in Equation~\ref{eq:sigAfit} assumes a top-hat filter in $k$ space; to convert it to a measured counts-in-cells variance $\sigma_A^2(\ell)$, one must account for both the mass-assignment function and alias effects. Equations~\ref{eq:P_waa} and \ref{eq:sigR} quote a prescription for doing so from \citet{Jing2005}.

The skewness $T_3$ of the $A$-distribution depends in turn on both the variance $\sigma_A^2(\ell)$ and the slope $n_s$ of the no-wiggle linear power spectrum at the smoothing scale $\ell$. Describing this dependence requires four free parameters; the description appears in Equations~\ref{eq:T3} through \ref{eq:T3params}.

Having expressions for the first three moments of $A$, we show that a generalized extreme value (GEV) distribution of the reversed Wiebull type matches the actual $A$-distribution well (Equations~\ref{eq:GEV} through \ref{eq:GEV_mu}). When we compare the cumulative distribution functions of the actual $A$-distribution to those generated by our GEV prescription, we find the maximum difference to be less than 2 per cent, with rms differences at most 0.8 per cent.

Having thus characterized the distribution of the log density field $A$, it becomes trivial to write the distribution for the actual density perturbation field $\delta$. However, it is the power spectrum of $A$, not of $\delta$, which captures the majority of the cosmological information at small scales \citep{CarronSzapudi2013} and which hence will allow us to make full use of the data from upcoming galaxy surveys.

\section*{Acknowledgements}
The Millennium Simulation data bases used in this work and the web application providing online access to them were constructed as part of the activities of the German Astrophysical Virtual Observatory (GAVO). IS acknowledges support from National Science Foundation (NSF) award 1616974. We also thank Mark Neyrinck for his comments and suggestions, which significantly strengthened this paper.

\bibliography{ModelLogDensity}

\begin{thebibliography}{}
\makeatletter
\relax
\def\mn@urlcharsother{\let\do\@makeother \do\$\do\&\do\#\do\^\do\_\do\%\do\~}
\def\mn@doi{\begingroup\mn@urlcharsother \@ifnextchar [ {\mn@doi@}
  {\mn@doi@[]}}
\def\mn@doi@[#1]#2{\def\@tempa{#1}\ifx\@tempa\@empty \href
  {http://dx.doi.org/#2} {doi:#2}\else \href {http://dx.doi.org/#2} {#1}\fi
  \endgroup}
\def\mn@eprint#1#2{\mn@eprint@#1:#2::\@nil}
\def\mn@eprint@arXiv#1{\href {http://arxiv.org/abs/#1} {{\tt arXiv:#1}}}
\def\mn@eprint@dblp#1{\href {http://dblp.uni-trier.de/rec/bibtex/#1.xml}
  {dblp:#1}}
\def\mn@eprint@#1:#2:#3:#4\@nil{\def\@tempa {#1}\def\@tempb {#2}\def\@tempc
  {#3}\ifx \@tempc \@empty \let \@tempc \@tempb \let \@tempb \@tempa \fi \ifx
  \@tempb \@empty \def\@tempb {arXiv}\fi \@ifundefined
  {mn@eprint@\@tempb}{\@tempb:\@tempc}{\expandafter \expandafter \csname
  mn@eprint@\@tempb\endcsname \expandafter{\@tempc}}}

\bibitem[\protect\citeauthoryear{{Angulo} \& {White}}{{Angulo} \&
  {White}}{2010}]{AnguloWhite}
{Angulo} R.~E.,  {White} S.~D.~M.,  2010, \mn@doi [\mnras]
  {10.1111/j.1365-2966.2010.16459.x}, \href
  {http://adsabs.harvard.edu/abs/2010MNRAS.405..143A} {405, 143}

\bibitem[\protect\citeauthoryear{{Antal}, {Sylos Labini}, {Vasilyev}  \&
  {Baryshev}}{{Antal} et~al.}{2009}]{Antal2009}
{Antal} T.,  {Sylos Labini} F.,  {Vasilyev} N.~L.,   {Baryshev} Y.~V.,  2009,
  \mn@doi [EPL (Europhysics Letters)] {10.1209/0295-5075/88/59001}, \href
  {http://adsabs.harvard.edu/abs/2009EL.....8859001A} {88, 59001}

\bibitem[\protect\citeauthoryear{{Bertin} \& {Clusel}}{{Bertin} \&
  {Clusel}}{2006}]{BertinClusel}
{Bertin} E.,  {Clusel} M.,  2006, \mn@doi [Journal of Physics A Mathematical
  General] {10.1088/0305-4470/39/24/001}, \href
  {http://adsabs.harvard.edu/abs/2006JPhA...39.7607B} {39, 7607}

\bibitem[\protect\citeauthoryear{{Carron}}{{Carron}}{2011}]{Carron2011}
{Carron} J.,  2011, \mn@doi [\apj] {10.1088/0004-637X/738/1/86}, \href
  {http://adsabs.harvard.edu/abs/2011ApJ...738...86C} {738, 86}

\bibitem[\protect\citeauthoryear{{Carron} \& {Neyrinck}}{{Carron} \&
  {Neyrinck}}{2012}]{CarronNeyrinck2012}
{Carron} J.,  {Neyrinck} M.~C.,  2012, \mn@doi [\apj]
  {10.1088/0004-637X/750/1/28}, \href
  {http://adsabs.harvard.edu/abs/2012ApJ...750...28C} {750, 28}

\bibitem[\protect\citeauthoryear{{Carron} \& {Szapudi}}{{Carron} \&
  {Szapudi}}{2013}]{CarronSzapudi2013}
{Carron} J.,  {Szapudi} I.,  2013, \mn@doi [\mnras] {10.1093/mnras/stt1215},
  \href {http://adsabs.harvard.edu/abs/2013MNRAS.434.2961C} {434, 2961}

\bibitem[\protect\citeauthoryear{{Carron} \& {Szapudi}}{{Carron} \&
  {Szapudi}}{2014}]{CarronSzapudi2014}
{Carron} J.,  {Szapudi} I.,  2014, \mn@doi [\mnras] {10.1093/mnrasl/slt167},
  \href {http://adsabs.harvard.edu/abs/2014MNRAS.439L..11C} {439, L11}

\bibitem[\protect\citeauthoryear{{Coles}}{{Coles}}{2001}]{GEVColes}
{Coles} S.,  2001, {An introduction to statistical modeling of extreme values}.
Springer-Verlag, London

\bibitem[\protect\citeauthoryear{{Coles} \& {Jones}}{{Coles} \&
  {Jones}}{1991}]{ColesJones}
{Coles} P.,  {Jones} B.,  1991, \mn@doi [\mnras] {10.1093/mnras/248.1.1}, \href
  {http://adsabs.harvard.edu/abs/1991MNRAS.248....1C} {248, 1}

\bibitem[\protect\citeauthoryear{{Colombi}}{{Colombi}}{1994}]{Colombi1994}
{Colombi} S.,  1994, \mn@doi [\apj] {10.1086/174834}, \href
  {http://adsabs.harvard.edu/abs/1994ApJ...435..536C} {435, 536}\\

\bibitem[\protect\citeauthoryear{{Eisenstein} \& {Hu}}{{Eisenstein} \&
  {Hu}}{1998}]{EisensteinHu}
{Eisenstein} D.~J.,  {Hu} W.,  1998, \mn@doi [\apj] {10.1086/305424}, \href
  {http://adsabs.harvard.edu/abs/1998ApJ...496..605E} {496, 605}

\bibitem[\protect\citeauthoryear{{Fry} \& {Peebles}}{{Fry} \&
  {Peebles}}{1978}]{FryPeebles1978}
{Fry} J.~N.,  {Peebles} P.~J.~E.,  1978, \mn@doi [\apj] {10.1086/156001}, \href
  {http://adsabs.harvard.edu/abs/1978ApJ...221...19F} {221, 19}

\bibitem[\protect\citeauthoryear{{Gazta\~{n}aga}}{{Gazta\~{n}aga}}{1994}]{Gaztanaga1994}
{Gazta\~{n}aga} E.,  1994, \mn@doi [\mnras] {10.1093/mnras/268.4.913}, \href
  {http://adsabs.harvard.edu/abs/1994MNRAS.268..913G} {268, 913}

\bibitem[\protect\citeauthoryear{{Gumbel}}{{Gumbel}}{1958}]{GEVGumbel}
{Gumbel} E.~J.,  1958, {Statistics of extremes}.
Columbia UP, New York

\bibitem[\protect\citeauthoryear{{Jing}}{{Jing}}{2005}]{Jing2005}
{Jing} Y.~P.,  2005, \mn@doi [\apj] {10.1086/427087}, \href
  {http://adsabs.harvard.edu/abs/2005ApJ...620..559J} {620, 559}

\bibitem[\protect\citeauthoryear{{Klypin}, {Prada}, {Betancort-Rijo}  \&
  {Albareti}}{{Klypin} et~al.}{2017}]{Klypin2017}
{Klypin} A.,  {Prada} F.,  {Betancort-Rijo} J.,   {Albareti} F.~D.,  2017,
  preprint, \href {http://adsabs.harvard.edu/abs/2017arXiv170601909K} {}
  (\mn@eprint {arXiv} {1706.01909})

\bibitem[\protect\citeauthoryear{{Leadbetter}, {Lindgren}  \&
  {Rootz\'{e}n}}{{Leadbetter} et~al.}{1983}]{GEVLB}
{Leadbetter} M.~R.,  {Lindgren} G.,   {Rootz\'{e}n} H.,  1983, {Statistics of
  extremes}.
Columbia UP, New York

\bibitem[\protect\citeauthoryear{{Lee}, {Primack}, {Behroozi},
  {Rodr{\'{\i}}guez-Puebla}, {Hellinger}  \& {Dekel}}{{Lee}
  et~al.}{2017}]{Lee2017}
{Lee} C.~T.,  {Primack} J.~R.,  {Behroozi} P.,  {Rodr{\'{\i}}guez-Puebla} A.,
  {Hellinger} D.,   {Dekel} A.,  2017, \mn@doi [\mnras]
  {10.1093/mnras/stw3348}, \href
  {http://adsabs.harvard.edu/abs/2017MNRAS.466.3834L} {466, 3834}

\bibitem[\protect\citeauthoryear{{Lewis} \& {Challinor}}{{Lewis} \&
  {Challinor}}{2002}]{CAMB}
{Lewis} A.,  {Challinor} A.,  2002, \mn@doi [\prd]
  {10.1103/PhysRevD.66.023531}, \href
  {http://adsabs.harvard.edu/abs/2002PhRvD..66b3531L} {66, 023531}

\bibitem[\protect\citeauthoryear{{Neyrinck}}{{Neyrinck}}{2013}]{Neyrinck2013}
{Neyrinck} M.~C.,  2013, \mn@doi [\mnras] {10.1093/mnras/sts027}, \href
  {http://adsabs.harvard.edu/abs/2013MNRAS.428..141N} {428, 141}

\bibitem[\protect\citeauthoryear{{Neyrinck} \& {Szapudi}}{{Neyrinck} \&
  {Szapudi}}{2007}]{NeyrinckSzapudi2007}
{Neyrinck} M.~C.,  {Szapudi} I.,  2007, \mn@doi [\mnras]
  {10.1111/j.1745-3933.2006.00275.x}, \href
  {http://adsabs.harvard.edu/abs/2007MNRAS.375L..51N} {375, L51}

\bibitem[\protect\citeauthoryear{{Neyrinck}, {Szapudi}  \& {Szalay}}{{Neyrinck}
  et~al.}{2009}]{NSS09}
{Neyrinck} M.~C.,  {Szapudi} I.,   {Szalay} A.~S.,  2009, \mn@doi [\apjl]
  {10.1088/0004-637X/698/2/L90}, \href
  {http://adsabs.harvard.edu/abs/2009ApJ...698L..90N} {698, L90}

\bibitem[\protect\citeauthoryear{{Repp} \& {Szapudi}}{{Repp} \&
  {Szapudi}}{2017}]{Repp2017}
{Repp} A.,  {Szapudi} I.,  2017, \mn@doi [\mnras] {10.1093/mnrasl/slw178},
  \href {http://adsabs.harvard.edu/abs/2017MNRAS.464L..21R} {464, L21}

\bibitem[\protect\citeauthoryear{{Repp}, {Szapudi}, {Carron}  \& {Wolk}}{{Repp}
  et~al.}{2015}]{Repp2015}
{Repp} A.,  {Szapudi} I.,  {Carron} J.,   {Wolk} M.,  2015, \mn@doi [\mnras]
  {10.1093/mnras/stv2212}, \href
  {http://adsabs.harvard.edu/abs/2015MNRAS.454.3533R} {454, 3533}

\bibitem[\protect\citeauthoryear{{Rimes} \& {Hamilton}}{{Rimes} \&
  {Hamilton}}{2005}]{RimesHamilton2005}
{Rimes} C.~D.,  {Hamilton} A.~J.~S.,  2005, \mn@doi [\mnras]
  {10.1111/j.1745-3933.2005.00051.x}, \href
  {http://adsabs.harvard.edu/abs/2005MNRAS.360L..82R} {360, L82}

\bibitem[\protect\citeauthoryear{{Shin}, {Kim}, {Pichon}, {Jeong}  \&
  {Park}}{{Shin} et~al.}{2017}]{Shin2017}
{Shin} J.,  {Kim} J.,  {Pichon} C.,  {Jeong} D.,   {Park} C.,  2017, \mn@doi
  [\apj] {10.3847/1538-4357/aa74b9}, \href
  {http://adsabs.harvard.edu/abs/2017ApJ...843...73S} {843, 73}

\bibitem[\protect\citeauthoryear{{Smith}, {Peacock}, {Jenkins}, {White},
  {Frenk}, {Pearce}, {Thomas}  et~al.}{{Smith} et~al.}{2003}]{Smith_et_al}
{Smith} R.~E.,  {Peacock} J.~A.,  {Jenkins} A.,  {White} S.~D.~M.,  {Frenk}
  C.~S.,  {Pearce} F.~R.,  {Thomas} P.~A.,   et~al., 2003, \mn@doi [\mnras]
  {10.1046/j.1365-8711.2003.06503.x}, \href
  {http://adsabs.harvard.edu/abs/2003MNRAS.341.1311S} {341, 1311}

\bibitem[\protect\citeauthoryear{{Springel} et~al.,}{{Springel}
  et~al.}{2005}]{Springel2005}
{Springel} V.,  et~al., 2005, \mn@doi [\nat] {10.1038/nature03597}, \href
  {http://adsabs.harvard.edu/abs/2005Natur.435..629S} {435, 629}

\bibitem[\protect\citeauthoryear{{Szapudi} \& {Pan}}{{Szapudi} \&
  {Pan}}{2004}]{SzapudiPan2004}
{Szapudi} I.,  {Pan} J.,  2004, \mn@doi [\apj] {10.1086/380920}, \href
  {http://adsabs.harvard.edu/abs/2004ApJ...602...26S} {602, 26}

\bibitem[\protect\citeauthoryear{{Szapudi}, {Szalay}  \& {Boschan}}{{Szapudi}
  et~al.}{1992}]{SSB1992}
{Szapudi} I.,  {Szalay} A.~S.,   {Boschan} P.,  1992, \mn@doi [\apj]
  {10.1086/171286}, \href {http://adsabs.harvard.edu/abs/1992ApJ...390..350S}
  {390, 350}

\bibitem[\protect\citeauthoryear{{Uhlemann}, {Codis}, {Pichon}, {Bernardeau}
  \& {Reimberg}}{{Uhlemann} et~al.}{2016}]{Uhlemann2016}
{Uhlemann} C.,  {Codis} S.,  {Pichon} C.,  {Bernardeau} F.,   {Reimberg} P.,
  2016, \mn@doi [\mnras] {10.1093/mnras/stw1074}, \href
  {http://adsabs.harvard.edu/abs/2016MNRAS.460.1529U} {460, 1529}

\bibitem[\protect\citeauthoryear{{Wang}, {Percival}, {Cimatti}, {Mukherjee},
  {Guzzo}, {Baugh}, {Carbone}  et~al.}{{Wang} et~al.}{2010}]{Wang2010}
{Wang} Y.,  {Percival} W.,  {Cimatti} A.,  {Mukherjee} P.,  {Guzzo} L.,
  {Baugh} C.~M.,  {Carbone} C.,   et~al., 2010, \mn@doi [\mnras]
  {10.1111/j.1365-2966.2010.17335.x}, \href
  {http://adsabs.harvard.edu/abs/2010MNRAS.409..737W} {409, 737}

\bibitem[\protect\citeauthoryear{{Wolk}, {McCracken}, {Colombi}, {Fry},
  {Kilbinger}, {Hudelot}, {Mellier}  \& {Ilbert}}{{Wolk}
  et~al.}{2013}]{Wolk2013}
{Wolk} M.,  {McCracken} H.~J.,  {Colombi} S.,  {Fry} J.~N.,  {Kilbinger} M.,
  {Hudelot} P.,  {Mellier} Y.,   {Ilbert} O.,  2013, \mn@doi [\mnras]
  {10.1093/mnras/stt1111}, \href
  {http://adsabs.harvard.edu/abs/2013MNRAS.435....2W} {435, 2}

\bibitem[\protect\citeauthoryear{{Wolk}, {Carron}  \& {Szapudi}}{{Wolk}
  et~al.}{2015}]{WCS2015}
{Wolk} M.,  {Carron} J.,   {Szapudi} I.,  2015, \mn@doi [\mnras]
  {10.1093/mnras/stv1891}, \href
  {http://adsabs.harvard.edu/abs/2015MNRAS.454..560W} {454, 560}

\makeatother
\end{thebibliography}

\section*{Appendix}
We here briefly demonstrate that both \citet{Uhlemann2016} and our GEV model predict double-exponential behavior.

We begin with equations~5, 12, and 13 of \citeauthor{Uhlemann2016}, which give us
\begin{eqnarray}
\Psi_R(\hat{\rho}) & = & \frac{1}{2\sigma^2(r)}\tau(\hat{\rho})^2,\hspace{0.5cm}r^3=R^3 \hat{\rho}\\
\hat{\rho}(\tau) & \approx & \frac{1}{\left(1 - \tau/\nu\right)^\nu}\\
\sigma^2(R) & = & \sigma^2(R_p)\left(R/R_p\right)^{-(n_s+3)}.
\end{eqnarray}
Note that $\hat{\rho}$ is the normalized density; $\hat{\rho} = \rho/\overline{\rho} = 1+\delta$ in our notation. From these expressions we obtain
\begin{equation}
\Psi_R(\hat{\rho}) = \frac{\nu^2}{2\sigma^2(R_p)}\left(\frac{R}{R_p}\right)^{n_s+3}\left( \hat{\rho}^{\left(\frac{n_s+3}{6}\right)} - \hat{\rho}^{\left(\frac{n_s+3}{6}-\frac{1}{\nu}\right)} \right)^2.
\end{equation}
From their equation 11 we can write
\begin{equation}
\mathcal{P}(A) = \hat{\rho}\left[\frac{\Psi_R''(\hat{\rho}) + \Psi_R'(\hat{\rho})/\hat{\rho}}{2\pi}\right]^{1/2} \exp\left(-\Psi_R(\hat{\rho})\right),
\end{equation}
so that after some algebra,
\begin{eqnarray}
\lefteqn{\mathcal{P}(A)=\frac{\nu}{\sigma(R_p)\sqrt{2\pi}}\left(\frac{R}{R_p}\right)^{\frac{n_s+3}{2}}
\hat{\rho}^{\left(\frac{n_s+3}{6}\right)}}\label{eq:long}\\
& & \nonumber\times \left\{\frac{(n_s+3)^2}{18} - \left(\frac{n_s+3}{3} - \frac{1}{\nu}\right)^2 \hat{\rho}^{-1/\nu}\right.\\
& & \nonumber\hspace{1cm}+\left.2\left(\frac{n_s+3}{6} - \frac{1}{\nu}\right)^2 \hat{\rho}^{-2/\nu} \right\}^{1/2}\\
& & \nonumber\times \exp \left[ \frac{-\nu^2}{2\sigma^2(R_p)}\left(\frac{R}{R_p}\right)^{n_s+3} \hat{\rho}^{\left(\frac{n_s+3}{3}\right)} \left(1 - \hat{\rho}^{-1/\nu} \right)^2 \right].\\
& & \nonumber
\end{eqnarray}

Since by definition $\hat{\rho} = e^A$, the final line of the above equation (our Equation~\ref{eq:long}) will be a double exponential.

For the GEV distribution, it is clear that the function $t(A)$ in Equation~\ref{eq:GEV_t} becomes exponential for low values of $\xi$.  For the cases we consider, the values of $\xi$ are typically quite small, causing double-exponential behavior in our Equation~\ref{eq:GEV}.

\label{lastpage}
\end{document}